\documentclass[twocolumn,showpacs,preprintnumbers,amsmath,amssymb,prb]{revtex4}
\usepackage{graphicx}
\usepackage{dcolumn} 
\usepackage{bm} 
\usepackage[hypertex]{hyperref}
\usepackage{longtable}

\begin{document}

\title{Interference of spin states in photoemission from Sb/Ag(111)}
\author{Fabian Meier$^{1,2}$}
\author{Vladimir Petrov$^{3}$}
\author{Hossein Mirhosseini$^{4}$}
\author{Luc Patthey$^{2}$}
\author{J\"urgen Henk$^{4}$}
\author{J\"urg Osterwalder$^{1}$}
\author{J. Hugo Dil$^{1,2}$}
\affiliation{
$^{1}$Physik-Institut, Universit\"at Z\"urich, Winterthurerstrasse 190, 
CH-8057 Z\"urich, Switzerland 
\\ 
$^{2}$ Swiss Light Source, Paul Scherrer Institut, CH-5232 Villigen, 
Switzerland 
\\ 
$^{3}$St. Petersburg Polytechnical University, 29 Polytechnicheskaya St, 
195251 St Petersburg, Russia
\\
$^{4}$Max-Planck-Institut f\"ur Mikrostrukturphysik, D-06120 Halle (Saale), Germany}

\date{\today}

\begin{abstract}
Using a three-dimensional spin polarimeter we have gathered evidence for the interference of spin states in photoemission from the surface alloy Sb/Ag(111). This system features a small Rashba-type spin-splitting of a size comparable to the linewidth of the quasiparticles, thus causing an intrinsic overlap between states with orthogonal spinors. Besides a small spin polarization caused by the spin-splitting, we observe a large spin polarization component in the plane normal to the quantization axis provided by the Rashba effect. Strongly suggestive of coherent spin rotation, this effect is largely independent of the photon energy and photon polarization.
\end{abstract}


\maketitle
An important branch of spintronics research is looking for new systems with naturally existing spin polarized electrons and ways to manipulate their spins. The broken spacial inversion symmetry at surfaces can induce a spin splitting of electronic states in non-magnetic systems via the spin-orbit interaction. A substantial splitting due to this so-called Rashba effect~\cite{rashba:1984} was observed for the Shockley surface state on Au(111) by angle-resolved photoemission spectroscopy (ARPES)~\cite{LaShell:1996}. Later spin-resolved experiments confirmed the high degree of spin polarization of the electrons photoemitted from these states~\cite{Hoesch:2004}, observing helical spin structures tangential to the two spin split Fermi surfaces. More recently, surface alloys of Bi and Pb on Ag(111) have attracted much attention in the search for even larger spin splittings, exploiting a combination of strong atomic spin-orbit interaction of the heavy metals with structural effects enhancing the local potential gradients at the surface~\cite{Ast:2007, Pacile:2006}.

In this Letter we discuss the structurally related system of Sb on Ag(111) which has a small but finite spin splitting~\cite{moreschini:2009}. The splitting is so small that it cannot be resolved by ARPES in most of the surface Brillouin zone. Our spin-polarized ARPES data show nevertheless substantial spin polarization and permit to quantify the spin splitting. More importantly, the measured spin texture is at strong variance with that expected from the Rashba model and suggests that coherent superposition of spin states occur in the photocurrent. It is an intriguing property of quantum mechanics that spin states can interfere, hence the expectation value of the sum of two spinors can differ from the sum of the individual expectation values. In particular, the addition of a spin-up and a spin-down spinor along some quantization axis does not yield zero polarization, but results in a spinor with an expectation value (henceforth spin polarization) placed within the plane orthogonal to the quantization axis. This is exactly what we observe.

\begin{figure}[t]
\begin{center}
\includegraphics[width=0.4\textwidth]{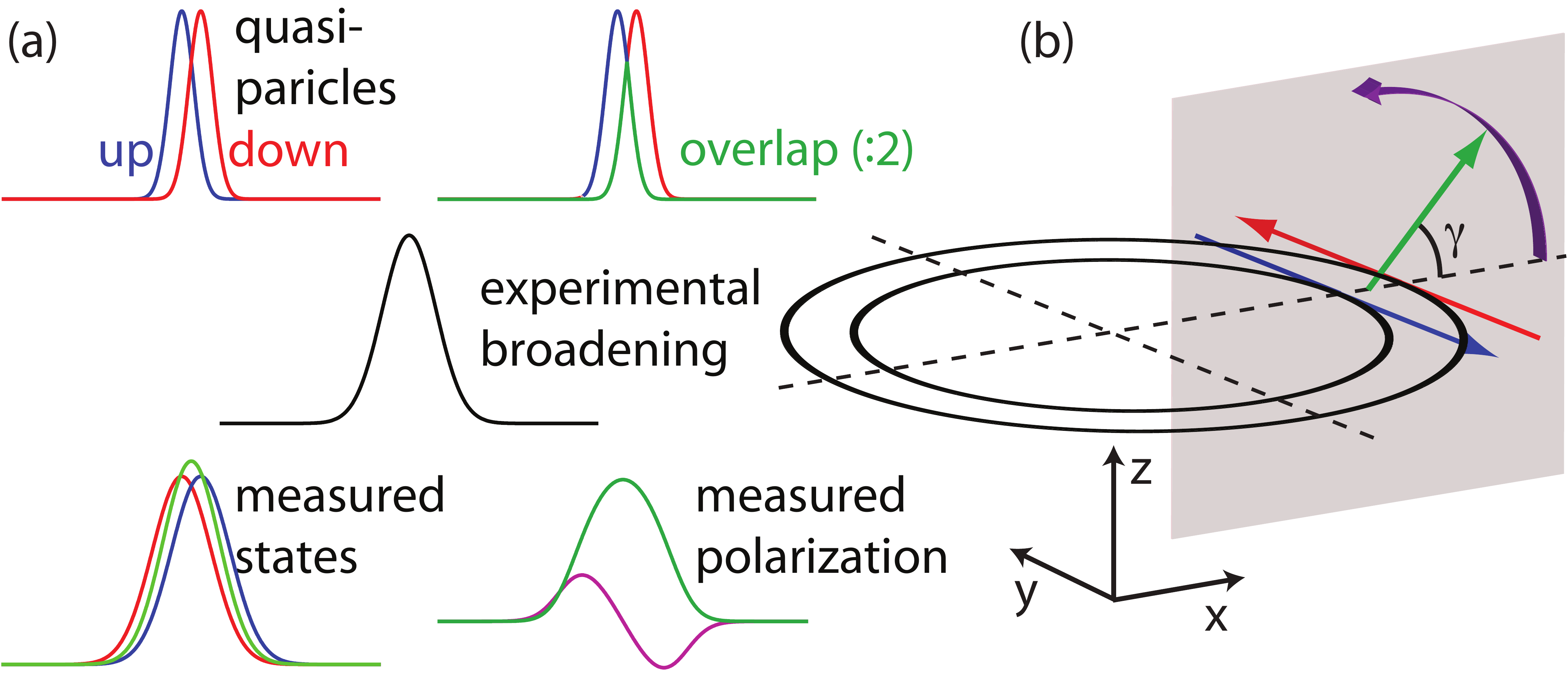}
\caption{
(color online) (a) and (b) Sketch of the suggested mechanism leading to spin polarization in the $xz$-plane by spin-state interference. (a) Small spin splitting with intrinsic overlap between the spin-up and spin-down quasiparticle excitations. A convolution with the experimental broadening leads to the strongly overlapping measured spectra, a Rashba-type spin polarization along the $y$ direction (violet/dark gray curve), and a coherently rotated spin polarization within the $xz$-plane (green/light gray curve). (b) Illustration of the associated spin polarization vectors in one region of the spin-split circular Fermi surface.} 
\label{Fig1}
\end{center}
\end{figure}

Spin-state interference has previously been observed in resonant photoemission induced by circularly polarized light from magnetized Gd by M\"uller et al.~\cite{muller:161401}. In this system, orthogonal spin states can be prepared by the angular momentum transfer from the light and spin-orbit interaction on one hand, and by direct photoemission from magnetized states in the valence band on the other hand. By tuning the photon energy to the 4d resonance, the two spin states can be brought to interfere. In our case it is the Rashba effect that defines the two orthogonal spin states, and we argue that they interfere when their energy or momentum splitting is of the same order as the intrinsic line width of either state, as is illustrated in Fig.~\ref{Fig1}. In the overlapping region the quasiparticles formed by the photohole states in the two spin-split bands are indistinguishable in time and space, and the corresponding photoelectrons should thus represent coherent superpositions as reflected in spin space. 

Like in the related Bi and Pb surface alloys on Ag(111), the Sb adatoms replace every third Ag atom in the topmost layer to form a ($\sqrt{3}\times\sqrt{3})R30^{\circ}$ superstructure~\cite{deVries1998159, woodruff:2000}, henceforth termed Sb/Ag(111). Mixed Sb$_{1-x}$Bi$_x$/Ag(111) layers, where Sb is randomly substituted by Bi, were also investigated. In such mixed alloys the spin splitting can be enhanced \cite{Ast:2008, Meier:2009PRB}, and they can therefore serve as a test for our overlap hypothesis. Finally, photoemission experiments were carried out with different photon energies and photon polarization in order to probe the dependence of the spin state interference on these parameters. 

\begin{figure}[htb]
\begin{center}
\includegraphics[width=0.4\textwidth]{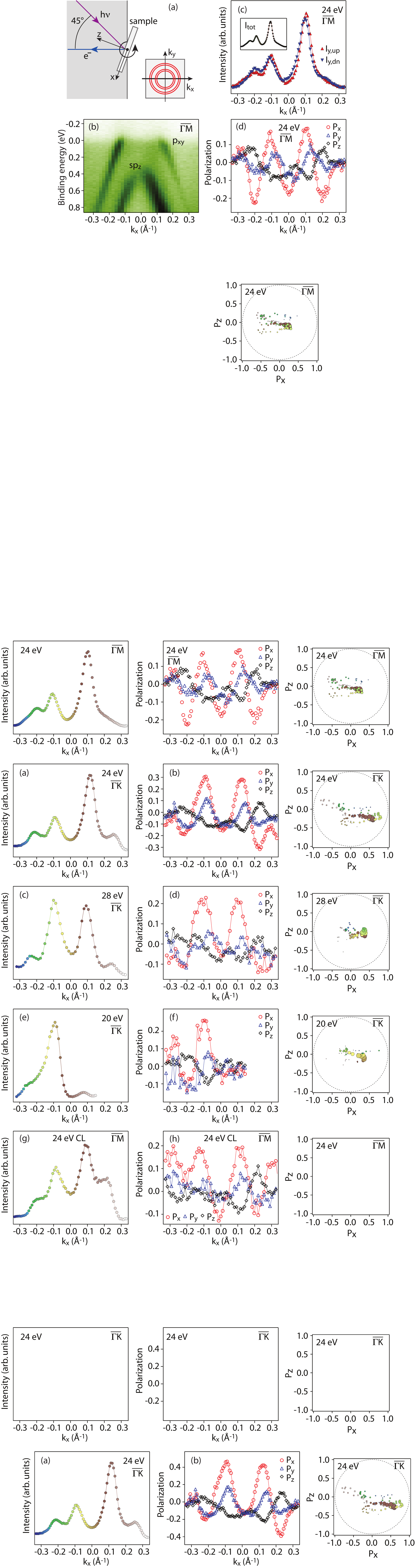}
\caption{
(color online) (a) Schematic experimental setup. (b) Spin integrated surface state band structure of the Sb/Ag(111) surface alloy around $\bar{\Gamma}$. (c) Spin resolved and spin integrated (inset) MDC intensity data for 24 eV photons in the direction $\bar{\Gamma}\bar{M}$ at $E_b=0.6$ eV. (d) Simultaneously obtained spin polarization curves for all three components.
        }
\label{Fig2}
\end{center}
\end{figure}

The spin-polarized ARPES (SARPES) experiments were performed at room temperature at the Surface and Interface Spectroscopy beamline at the Swiss Light Source of the Paul Scherrer Institute using the COPHEE spectrometer.~\cite{Hoesch:2002} The energy and angle resolution when measured with the Mott detectors was 80 meV and $\pm$ 0.75$^{\circ}$, respectively. The photoemission setup is schematically shown in Fig.~\ref{Fig2}~(a). There is a 45$^{\circ}$ angle between the incoming photons and the detected electrons. The $z$ axis is given by the sample normal and the sample is rotated around the $y$ axis. In a momentum distribution curve (MDC) this corresponds to a scan along the $k_x$ axis with $k_y=0$ as shown in Fig.~\ref{Fig2}~(a) for schematically drawn circular constant energy surfaces.

Sample preparation was carried out \textit{in situ} under ultrahigh vacuum conditions with a base pressure better than $2\times10^{-10}$ mbar. The Ag(111) crystal was cleaned by multiple cycles of Ar$^{+}$ sputtering and annealing. A total amount of Sb corresponding to 1/3 of a monolayer was evaporated from a calibrated evaporator at a pressure below $4\times10^{-10}$ mbar followed by an annealing at $T\approx300^{\circ}$ C. The sample quality was confirmed by low-energy electron diffraction, which showed sharp $\sqrt{3}\times\sqrt{3}$-spots, and ARPES. For the mixed alloys the deposition of Sb and Bi was simultaneous with a total amount of 1/3 monolayer.

In Fig.~\ref{Fig2}~(b) we show the spin integrated surface state band dispersion of the Sb/Ag(111) surface alloy around $\bar{\Gamma}$ measured along $\bar{\Gamma}\bar{M}$. Similar to the two related surface alloys Bi/Ag(111) and Pb/Ag(111)~\cite{Ast:2007, Pacile:2006, Meier:2008}, two sets of bands are observed, termed in analogy as $sp_z$ and $p_{xy}$. However, the Rashba-type spin splitting is here much smaller and is not resolved in the data of Fig.~\ref{Fig2}~(b). The smaller splitting can be understood as a consequence of the smaller atomic number $Z$ of Sb ($Z=51$) compared to Bi (83) and Pb (82) and a smaller surface corrugation~\cite{moreschini:2009, Bihlmayer:2007}. 

Bi/Ag(111) and Pb/Ag(111) show Rashba-type spin structures~\cite{Meier:2008}, i.e. the spin polarization is mainly in plane and orthogonal to the electron momentum ($P_y$ component).  In Figs.~\ref{Fig2}~(d) we give SARPES data obtained for Sb/Ag(111) at $E_b=0.6$ eV with $p$-polarized photons of 24 eV (i.e. light polarization in the $xz$ plane). Here, the spin polarization component $P_x$ is dominant, corresponding to the radial direction of the constant energy surfaces. It shows large modulation amplitudes \emph{centered} at the peak positions of the MDCs. This is in sharp contrast to the other systems~\cite{Dil:2009R} and represents a major deviation from the Rashba model. $P_y$ and $P_z$ components show modulations with smaller amplitudes, with those in $P_z$ being rather in phase with those in $P_x$. On the other hand, $P_y$ crosses the zero line at the peak centers, which is typical of Rashba-type behaviour~\cite{Meier:2008}.  From this latter curve we can produce the spin-resolved spectra as projected onto the $y$ axis~\cite{MeierNJP:2009}, which corresponds to the spin quantization axis in the Rashba model (Fig.~\ref{Fig2}~(c)). The spin-resolved MDCs $I_{y,\mathrm{up}}$ and $I_{y,\mathrm{dn}}$ show a clear signature of a Rashba-type spin-orbit splitting with a momentum shift $dk=2k_0\cong0.01$ \AA$^{-1}$\ between the two bands (as obtained from fitting the two main peaks). 

Another remarkable observation in these data is that the measured spin polarization curves violate time-reversal symmetry. According to this symmetry, the two spin-split partners of the Kramers pairs should have opposite spin polarization vectors for equivalent binding energies, i.e. $\boldsymbol{P}(\boldsymbol{k_{\parallel}})=-\boldsymbol{P}(-\boldsymbol{k_{\parallel}})$. Yet, the polarization curves $P_x$ and $P_z$ are symmetric with respect to $k_x=0$. The missing time-reversal symmetry is a strong indication of a photoemission related effect, since time-reversal symmetry has to hold for the quasiparticle wave functions. We suggest that the origin of this photoemission effect is the spin-state interference caused by the intrinsic overlap in each Kramers pair associated with the small spin splitting.

The coherent superposition of a spin-up and a spin-down spinor leads to a spin state with a spin polarization vector normal to the spin polarization of the individual spinors. For instance, an electron with its spin along the positive $z$ axis can be represented by two spinors with spins along the $y$ axis, reading $\sqrt{2}\cdot\left\langle z^{\uparrow} \right|=\left(\begin{matrix}1,1\end{matrix}\right)=\left(\begin{matrix}1,0\end{matrix}\right)+\left(\begin{matrix}0,1\end{matrix}\right)=\left\langle y^{\uparrow} \right|+\left\langle y^{\downarrow} \right|.$ Similarly, a spin along $x$ can be written as $\sqrt{2}\cdot\nolinebreak\left\langle x^{\uparrow} \right|=\left(\begin{matrix}1,-i\end{matrix}\right)=\left\langle y^{\uparrow} \right|+e^{i3\pi/2}\left\langle y^{\downarrow} \right|.$ A phase difference between $\left\langle y^{\uparrow} \right|$ and $\left\langle y^{\downarrow} \right|$ causes a rotation of the resulting spin polarization vector in the $xz$ plane.

The model illustrated in Figs.~\ref{Fig1}~(a) and (b) can now directly be applied to the case of  Sb/Ag(111). Here, the states are split by $2k_0\cong0.01$ \AA$^{-1}$\ and their spinors are well defined, termed up and down on the upper left of Fig.~\ref{Fig1}~(a). We were not able to measure the intrinsic line width of these peaks due to experimental limitations and the limited sample quality, but a realistic lower boundary is given by a value of 40 meV at room temperature, since for Au(111) at normal emission 21 meV were reported (at 30 K)~\cite{Reinert:2001}, and away from normal emission more than 40 meV were measured (at 60 K)~\cite{Nechaev:2009}. In the MDC, this value translates into an intrinsic momentum broadening of $0.01$ \AA$^{-1}$ when applying the group velocity measured at a binding energy of $0.6$ eV. There is thus clearly a large \emph{intrinsic} overlap between the two peaks. The peaks can then be divided in regions with purely spin up, purely spin down and an overlap region (upper right) in order to obtain the scenario shown on the lower left of Fig.~\ref{Fig1}~(a). Considering that the coherent addition of two orthogonal spinors has a spin polarization vector in the plane normal to the spin polarization of the initial spinors as shown in Fig.~\ref{Fig1}~(b), spin polarization curves like those on the lower right of Fig.~\ref{Fig1}~(a) are obtained. In particular, the spin polarization of the overlap region has components $P_x$ and $P_z$ with their maxima at the point of maximum overlap, i.e. centered on the MDC peak. The direction of the spin polarization in the $xz$ plane, described by the angle $\gamma$, is defined by the phase difference between the two orthogonal spin states of the Kramers pair. From the observation that the $P_x$ and $P_z$ curves are symmetric with respect to $k_x=0$ \AA$^{-1}$ (Fig. \ref{Fig2}~(d)), we conclude that corresponding states of opposite $k_x$ have equal spin rotation angles. For the $sp_z$ states we measure $\gamma = 22^\circ \pm 9^\circ$, for the $p_{xy}$ states the value is $-25 ^\circ \pm 10^\circ$. 

\begin{figure}[htb]
\begin{center}
\includegraphics[width=0.44\textwidth]{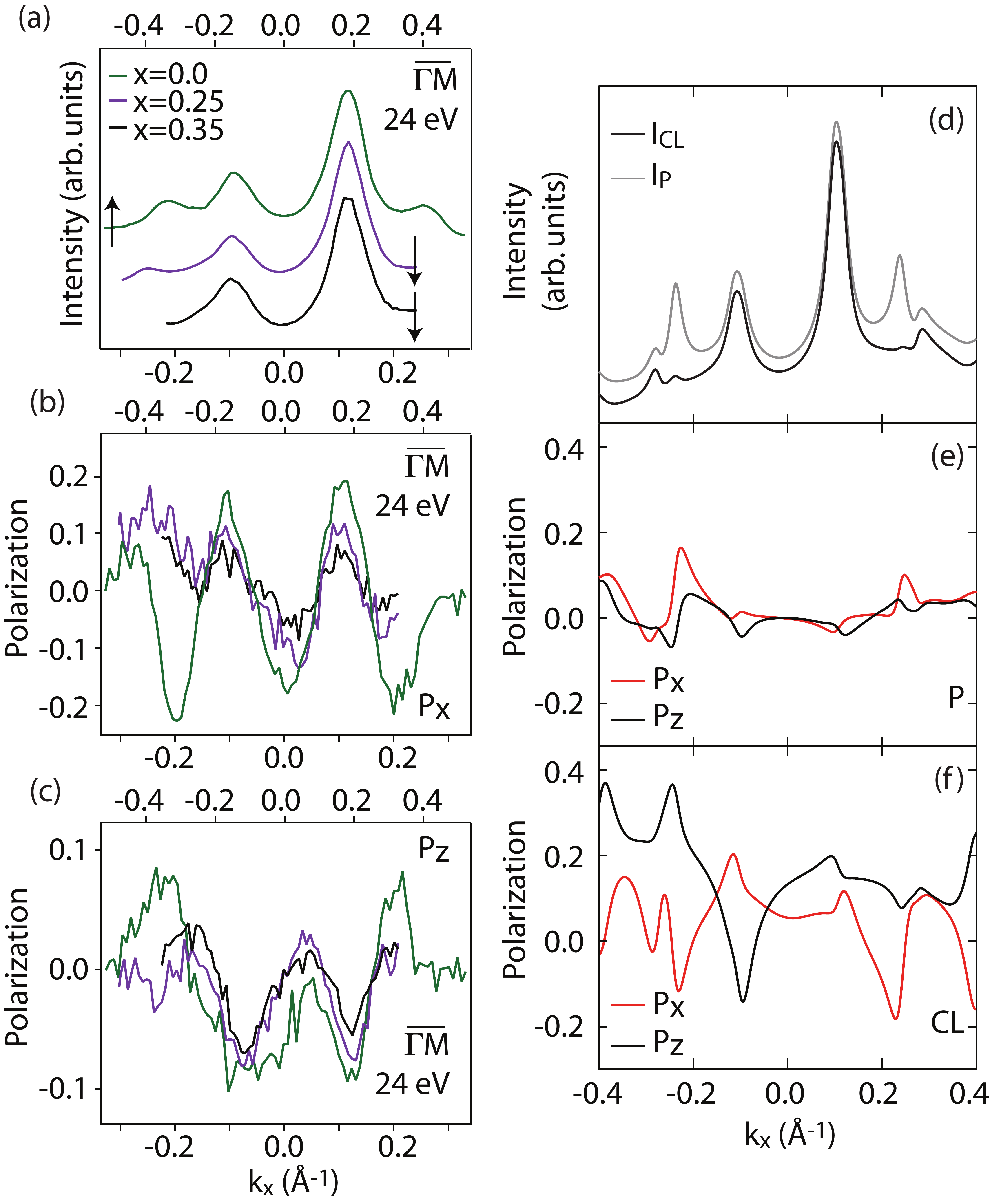}
\caption{
(color online) (a) Spin-integrated MDC data of Sb$_{1-x}$Bi$_{x}$/Ag(111) for $x=0$, $0.25$ and $0.35$ at $E_b=0.6$ eV ($x=1$) and 0.9 eV ($x=0.25$ and $0.35$). For better comparison, the $k_x$ scale for the $x=0$ sample (top curve) is given on the upper side of the frame, while the scale for $x=0.25$ and $0.35$ samples is given on the lower side . The arrows refer to the corresponding $k_x$ scale. (b) and (c) Spin polarization data $P_x$ and $P_z$ corresponding to the same samples as in (a). Note again the different $\boldsymbol{k}$-scales. (d)-(f) Fully relativistic spin-resolved one-step photoemission calculations for Sb/Ag(111) at $E_b=0.6$ eV and $hv=21.2$ eV. (d) The total (spin-integrated) photoemission intensity for $p$-polarized ($I_{P}$) and for circular left polarized light ($I_{CL}$). Spin polarization curves $P_x$ and $P_z$ for (e) $p$-polarized and (f) circular left polarized light.
        }
\label{Fig3}
\end{center}
\end{figure}

In Fig.~\ref{Fig3}~(a)-(c)  we show SARPES data for Sb$_{1-x}$Bi$_{x}$ for $x=0, 0.25$ and 0.35. From the spin polarization data $P_y$ (not shown) we find that the spin-splitting increases as Sb/Ag(111) is doped with Bi from $2k_0=0.01$ for $x=0$ to $2k_0=0.019$ \AA$^{-1}$ for $x=0.35$. The amplitudes of the spin polarization curve $P_x$ (and $P_z$) decrease markedly as the splitting is increased.  We thus observe a decrease in the measured spin polarization in the plane normal to the spin quantization axis of the quasiparticles as their splitting gets larger and the intrinsic overlap is reduced. This is fully in line with our model.

We performed fully relativistic spin-resolved one-step photoemission calculations for $x=0$, with $p$-polarized and circular left polarized light using the experimental geometry. Fig.~\ref{Fig3}~(d) shows the (spin-integrated) intensities for these two light polarizations. The two strong peaks nearest to $k_x = \pm 0.1$ \AA$^{-1}$ represent emission from $sp_z$ states, while the split peaks at $|{k_x}| >  0.2$ \AA$^{-1}$ are due to $p_{xy}$ emission. Fig.~\ref{Fig3}~(e) shows the spin polarization curves $P_x$ and $P_z$ for $p$-polarized light. While the curve for $P_y$ (not shown for better viewing clarity of $P_x$ and $P_z$) is in good agreement with the experimental data, there is neither quantitative nor qualitative agreement for the  polarization curves in the $xy$ plane (cf. Fig.~\ref{Fig2} (d)). The experimental spin polarization amplitudes are larger and the shapes of the curves are very different. A change of the photon polarization to circular left leads to drastic changes in the predicted spin polarization curves, shown in Fig.~\ref{Fig3}~(f). 

In general such calculations are in good agreement with the experimental data~\cite{Henk:2004}. The dramatic failure in the present case indicates that an important ingredient is missing in the theoretical description.  Specifically, the calculations do not capture coherent initial state effects as the quasiparticles are described by a non-local spectral density, which is an incoherent superposition of initial states. Hence, the disagreement between the data shown in Fig.~\ref{Fig2}~(d) and the curves in Fig.~\ref{Fig3}~(e) may also hint at a coherent effect in the initial states. 

\begin{figure}[htb]
\begin{center}
\includegraphics[width=0.484\textwidth]{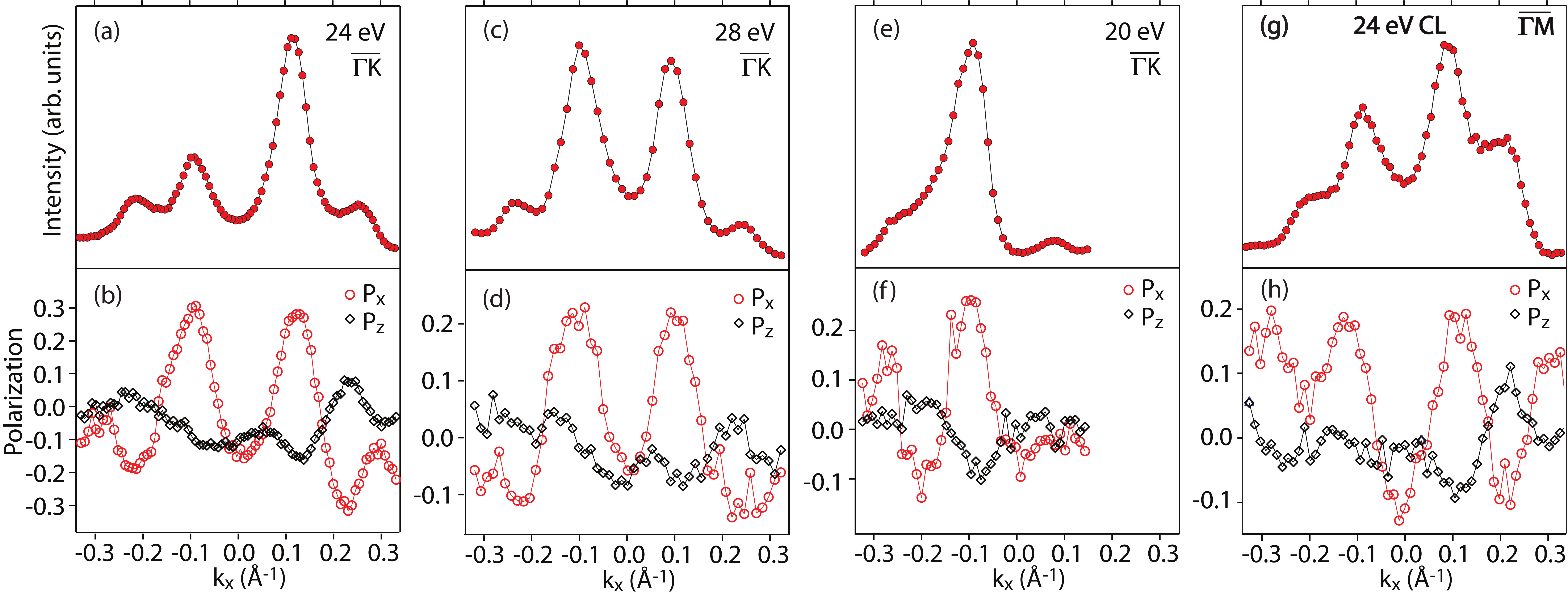}
\caption{
(color online) SARPES data for Sb/Ag(111) at $E_b=0.6$ eV for different photon energies and light polarization. (a), (c) and (e) MDC intensity data along $\bar{\Gamma}{K}$ for $p$-polarized light with photon energies of 24 eV, 28 eV and 20 eV, respectively. (b), (d) and (f) Corresponding spin polarization data. (g) MDC intensity data along $\bar{\Gamma}{M}$ for circular left polarized light with $h\nu=24$ eV. (h) Spin polarization data corresponding to (g). 
        }
\label{Fig4}
\end{center}
\end{figure}

Spin polarization observed in photoemission data can have various other origins \cite{Tamura:1987, Tamura:1994, Henk:1994, Schmiedeskamp:1988, Irmer:1988, Irmer:1996, Henk:2004}, but the outcome depends strongly on the symmetry of the solid and of the particular surface, on photon energy as well as on the absolute directions of photon incidence, photon polarization and electron emission. In order to rule out such effects, we have measured SARPES MDC data for Sb/Ag(111) at a binding energy of 0.6 eV for different photon energies and different light polarizations (Fig.~\ref{Fig4}). With respect to Fig.~\ref{Fig2}, the sample has been rotated by 90$^{\circ}$ and the MDCs are thus along $\bar{\Gamma}\bar{K}$ for Fig.~\ref{Fig4} (a)-(f), while (g) and (h) show again a scan along $\bar{\Gamma}\bar{M}$. The upper panels show the MDC intensity data, the lower ones the corresponding spin polarization curves. We observe that the effect is quite robust against variations of these experimental parameters. The sample rotation of 90$^{\circ}$ has no significant influence on the spin polarization curves (Fig.~\ref{Fig2} (d) vs. Fig.~\ref{Fig4} (b)), in contrast to the effects described by Tamura et al.~\cite{Tamura:1987} Second, although the intensity distribution curves change as a function of the photon energy, the spin polarization features are qualitatively not affected. The local extrema for $P_x$ ($P_z$) are always centered on the peaks, and are positive (negative) for the inner and negative (positive) for the outer ones. The absence of a photon energy dependence (Fig.~\ref{Fig4} (b), (d) and (f)) rules out a strong contribution of spin-orbit coupling in the final states. Most striking is the finding that a change from $p$-polarized to circular left polarized light (Fig.~\ref{Fig2} (d) vs. Fig.~\ref{Fig4} (h)) has no significant effect on the measured spin polarization curves, in contrast to the calculations shown in Fig. \ref{Fig3} (e) and (f). This corroborates the hypothesis that the measured $P_x$ and $P_z$ spin polarization is dominated by the spin structure of the initial state quasiparticles. 

Spin-state interference has recently been predicted for photoemission from the $\pi$ states of graphene~\cite{kuemmeth-2009}, where photoelectrons from equivalent atoms within the same unit cell interfere. The authors describe this effect as an interference between spin and pseudo-spin. In contrast, for Sb/Ag(111), the interference stems from an intrinsic overlap in $\boldsymbol{k}$ space.

In summary, we have presented evidence for a coherent superposition of spin states in photoemission from a Rashba system. Interference is assigned to the intrinsic overlap region of spin-up and spin-down states. Hence the measured spin polarization is defined by the quasiparticles in the initial state bands and is not significantly modified by final state effects. This was verified by varying the sample orientation, the photon energy and the light polarization. These observations are not described by relativistic one-step model calculations. What defines the phase difference between the states with opposite spin in the overlap region remains a key issue towards a further understanding of this effect.

Fruitful discussions with T. Greber, M. Hengsberger, B. Slomski, C. R. Ast and I. Gierz are gratefully acknowledged. 
We thank C. Hess, F. Dubi, and M. Kl\"ockner for technical support. This work is supported by the Swiss National Foundation.

\footnotesize
\bibliographystyle{apsrev}

\newpage

\end{document}